# Uberization of telecom networks for cost-efficient communication and computing

Hanna Bogucka[†‡], Bartosz Kopras[‡]

[†]Poznan University of Technology, [‡]RIMEDO Labs, Poland

**Abstract**

This paper discusses the uberization of telecommunication and computing network services. The Uber-like platform business model is discussed for application in future networks together with interesting analogies of communication and computing (2C) resource-sharing models. The economy of this sharing is discussed, and some recommendations for network uberization are provided.

**I. Introduction**

Ubiquitous internet access and innovations in communication and computing platforms explain new democratization trends in running a business. These trends are reflected in how different types of suppliers provide services and products through these platforms to many different types of buyers around the world. Uber is a good example of a digital platform-based service connecting consumers and occasional vendors in real-time. The Uber application, matching car owners with people who need rides, was launched in 2009 in the US. It has changed the competition rules of the economic game by allowing small players who possess means to provide services. Its success has inspired the term "uberization", which is seen as an important feature of economic transformation [1].

According to the Collins English Dictionary, "to uberize" means "*to subject (an industry) to a business model in which services are offered on demand through direct contact between a customer and a supplier, usually via mobile technology.*" In this model, service providers are freelancers with minimal contracts with the organization, flexible work conditions, and possessing and sharing the means of the service provision. Uberization is a part of sharing economy, a socio-economic system built around the sharing of resources [1]. It is now the modus operandi of many entrepreneurs worldwide in various branches of the economy.

Uberization has been considered an important possible trend in changing the telcos' business models for a few years [2]-[6] (e.g., [2] claims that a "*New revolution has just begun and telecom sector is about to be uberized with lower tariff plans and cost which is bare minimum*"). A very illustrative example is provided in a Forbes article [3], citing one of the uberized-market players: "*think AirBnB or Uber for telcos, if I have an empty room, power and internet why would I not leverage it to use it as a network tower.*" We could even go further to ask: *If I have a computer and internet, why would I not let it compute for my financial profit?* These are bold ideas with the prospect of open participation in the 2C business. However, the complexity of networks, diverse ownership of 2C resources, associated security issues, and administrative and market regulations (including spectrum regulations) may be viewed as limitations for this openness. The telecommunication service alone consists of i) networking technology, ii) operation support systems, iii) methods and procedures, and finally, iv) the content and applications [7]. In principle, all these elements can be uberized (as per the definition above), although not necessarily by a single service contractor. The same applies to computing services.

The motivation behind considering the uberization of communication and computing (2C) services is the following. First, breaking up existing oligopolies allows for open competition, transparency of operation, security scrutiny, and lower prices. Secondly, there is an increasing demand from the global business and individual customers for more and more computing power and resources providing it, while the existing (private or local) resources are underutilized. Sharing of 2C resources allows for more flexible and efficient utilization of these resources. This efficiency can be understood in different ways, e.g., energy efficiency, spectral efficiency, efficiency in delivering mission-critical services (with latency and bit-error-rate constraints), etc. For example, centralized computing services can be effective in computing energy costs, but the cost of communication to distanced servers and the associated delay may be unacceptable for mission-critical services. Thus, flexibility in offering and using appropriate resources for appropriate services should allow for the cost-effective delivery of services.

When considering uberization of the 2C services, it is important to undertake the challenge of proving the concept. Note that the proof of economic concept focuses on whether the business idea is viable and analyses the potential of the idea but does not test the demand it has in the market. This is why here, below, we structurally analyze how the platform (Uber-like) business model and its components can be adapted to this market. We try to answer whether uberization is a relevant concept for telecommunication and computing services and the IT world in general. To this end, we consider jointly technical and economic perspectives on shared 2C resources.

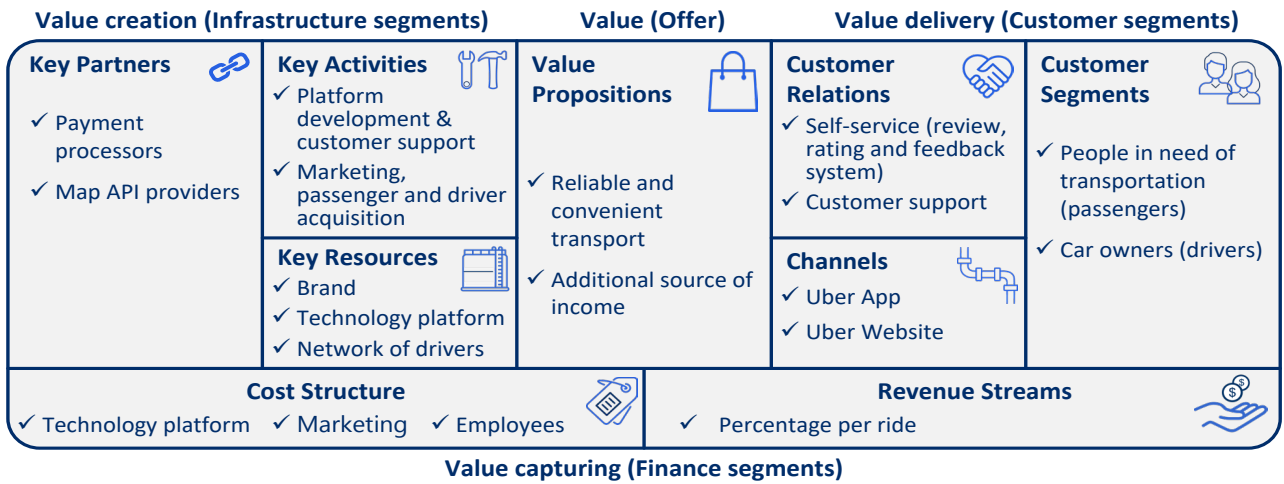

Fig. 1. Uber Business Model Canvas

The paper is structured as follows. The next section discusses the conceptual framework of a platform-business model to analyze whether uberization is a valid model for future 2C networks. The considerations of the economy of shared commodities and resources follow this. The final section suggests directions for further research and concludes our considerations.

## II. Uber business model for communication and computing

According to the Deloitte report [4], which discusses the future of the telecommunication landscape in 2030, uberization and sharing economy will be key drivers for the future of the telecommunication industry with a low degree of uncertainty and very high possible impact. Moreover, four future business-model scenarios have been identified resulting from two axes reflecting telco business uncertainties for 2030: *Ownership of the technology layer* and the *Dominance of the traditional customer relationship*. Telco uberization can be found on these axes pointing to the direction of technology competence remaining with telcos (although this may be gradually changing in the opposite direction) and telcos displaced from customer relationships (replaced by the customer-producer matching platform).

In its simplest form, a business model for a company offering some value can be broken down into three parts: i) Everything it takes to create this value (activities, resources, partners), ii) Everything it takes to deliver value to customers (customer relations, customer segments, channels), and iii) Everything that it takes to capture value (pricing strategy, payment methods, etc.). A universal definition and a visual framework to describe business models is Business Model Canvas which specifies these parts and elements (Fig. 1).

Article [5] discusses the telecos and Uber-like business models and their differences. Interestingly, telcos are adhering to a classical business model defined by a causal and sequential value chain, having a high degree of end-to-end customer experience control (thus, placed on the other extreme of the mentioned Deloitte *Dominance of the traditional customer relationship* axis, namely customer relationship remaining with telcos).

### A. Platform (Uber-like) Business Model

Unlike contemporary telecommunication companies, Uber, Airbnb, Booking.com, etc., have platform-based business models defined by a unique value proposition, which is to facilitate the connection of providers and consumers. They reclaim under-utilized assets owned by individuals (e.g., cars). Thus, a platform-based business relies on the sharing economy, monetizing networking consumers and producers. As such, platform businesses rarely have complete end-to-end customer experience control but focus on the quality and experience of networked connectivity.

Note that the platform-based business model is sometimes referred to as the *marketplace* business model. The marketplace is the place where customers and service providers meet directly without intercessors. The platform business model underpins the most successful companies, either fully or to a considerable degree. It includes the biggest market caps and many startups. The Uber business model canvas based on information from businessmodelanalyst.com is presented in Fig. 1.

A typical platform business's technology stack, implemented in a cloud is presented in Fig. 2. Its functionalities are residing on three layers: the *Networked Marketplace* that matches consumers with producers, using the digital (usually mobile) application, the *Enabling Layer* that includes enabling services, software tools, business logic, policies, and rules, and the data-driven decision making layer, namely *(Big) Data Layer* supported by Artificial Intelligence (AI) or Machine Learning (ML) algorithms [5]. These layers abstract from lower OSI layers: physical, data link control, network, or transport layers. Their operation is supported by the cloud infrastructure, including servers, databases, and telecommunication networks. Networking, however, is usually taken care of by telecommunication services and network providers. That is why platform businesses appear undemanding in infrastructure and capital/assets compared to telco providers, although very complex regarding software. The telco business model requires expensive infrastructure and assets such as operating licenses, public phone numbers, spectrum frequencies, etc. [6]. Thus, it seems that a prerequisite of transformation to a platform business for a telco operator is to become a virtual operator entering the business with light investments.

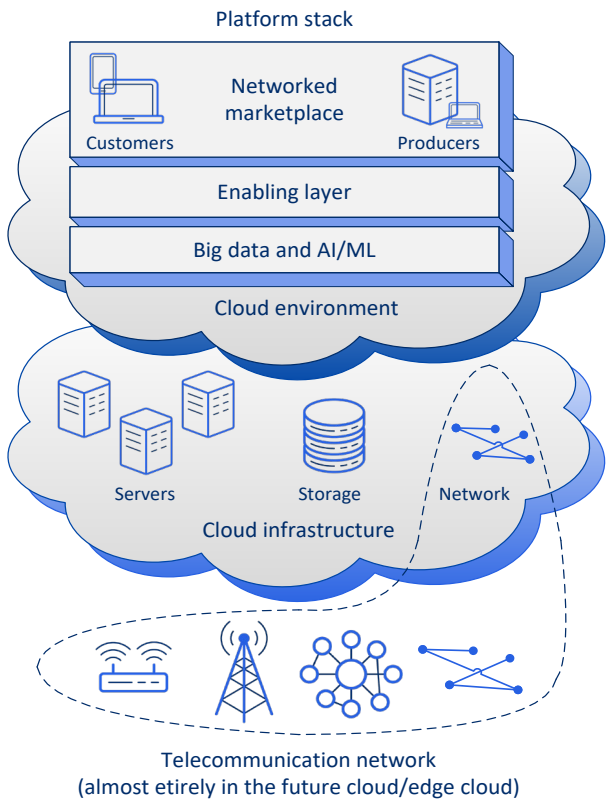

Fig. 2. A platform-business's technology stack, residing in a cloud.

Above, we have presented the existing Uberization concept, business model, and the platform business's technology stack. In what follows, we analyze the application of this business model (infrastructure, offer, customer, and finance segments) for the future 2C networks and services, compare Uber and Uber-like 2C networks operating business models to find analogies, discuss constraints related to technologies, Quality of Service (QoS) provision, regulations, spectrum policies, security and pricing mechanisms, and suggest the roadmap to implementation of open 2C market, which are our original contributions to the development of the networks uberization concept.

*B. Uberization options for telcos and 2C*

Telecommunication network has typically not been a part of the platform stack, although cloud infrastructure requires networking services, which are typically provided by telcos. However, it is very important to note that telecommunication networks and services are getting more virtualized in 5G and prospective 6G communication. Apart from core network functions virtualization, which is factual in 5G, open radio-access network (O-RAN) architecture has already been recommended [8] with software applications (xApps and rApps) for non-real-time and near-real-time Radio Intelligent Controller (RIC). Note that the O-RAN Alliance has been formed in 2021 by successful players in the telecom market and gathers more than 300 mobile operators, vendors, and research and academic institutions with the mission to reshape RANs to be more intelligent, open, virtualized, and fully interoperable. Research contributions and field trials allow considering O-RAN for practical implementations in the near future. Moreover, Multi-User Edge Computing (MEC) module is a part of 5G RAN bringing the cloud capabilities closer to the edge of the network, which is particularly important for mission-critical applications.

Virtualization of networks, and RAN functions together with edge computing and edge AI/ML allow considering the *Telecommunication Network* area from Fig. 2 to become another cloud (or edge cloud) infrastructure. This, together with operators' virtualization, opens the way for the telco business to transform into a platform business.

Contemporary IT services involve both communication and computing of information across the network, and thus, 2C services should be handled jointly. The computing (processing and storage) services are already available commercially as platform-based cloud services. Thus, incorporation of these services in uberized 2C platform-model-based business seems to be straightforward. The considered platform model should encompass under-utilized 2C resources such as computing machines (servers or less-powerful personal computers) and communication means (virtualized networks or private infrastructures).

Apart from sharing the infrastructure, radio spectrum resource sharing should be included in the considered model. Regarding mobile services, the traditional approach of regulatory bodies is the long-term spectrum licensing to operators, winning the spectrum auctions. However, spectrum sharing is considered an opportunity to open up access to a new spectrum for mobile services, and regulators need to help incentivize incumbents in attractive bands to share [9]. Citizens Broadband Radio Service (CBRS)-type approaches: the *Prioritised Access Licence* and *General Authorised Access* for the 3.5 GHz band in the USA, *License-*

Table I. Uber-analogies for uberized communication and computing network services

| | Car services - Uber | Network uberization |
|---|---|---|
| **Value proposition** | Car ride | Communication and computing, data processing and storage |
| **Shared resources** | Cars | Computing machines, databases, software, hardware |
| **Required infrastructure** | Roads | Telecom infrastructure, spectrum |
| **Customer segment – Service providers** | Car owners | Servers/cloud/edge cloud/computer/software owners, network infrastructure owners |
| **Customer segment - consumers** | Individual persons | IoT devices, companies, individuals |
| **Customer relations – quality assurance** | Customer review, rating, and feedback system | QoS metrics should be guaranteed by the service provider and monitored by the customer |
| **Technological platform - Matching platform** | Internet, Uber App | Internet, prospective App offering QoS matched, cost-optimized services |
| **Authentication/verification of service providers** | Driving license, compulsory technical inspection of a car | TBD, e.g., "Computing compliance license," homologation of equipment, blockchain-based smart contracts |
| **Authentication/verification of consumers** | Phone numbers, the customers' location | TBD, e.g., IP addresses, block-chain based smart contracts |
| **Security means** | Verification of drivers at enrolment, Real-Time ID check | Continuous verification, firewalls, authentication of service providers and consumers, "zero trust" philosophy |
| **Competition basis** | Service time and quality, the reputation of a driver | Service time and quality, the reputation of a server/service provider |
| **Regulations, related legal operating principles** | Traffic law, road/passenger transport legal acts | Legal acts of provision of electronic/telecommunication services, spectrum, and spectrum-sharing regulations |

*Shared Access* and *Authorized Shared Access* for the 2.3 GHz band in Europe, and *Concurrent Shared Access* allowed in 10 countries are well-established concepts that allow spectrum to be used by more than one operator. Moreover, spectrum sharing in Digital TV (DTV) white spaces (locally unused DTV channels) has already been validated by many field trials worldwide. Wireless access to the Internet is also possible in the shared ISM band, assuring high QoS in WiFi 6 standard. Contemporary radios can operate in multiple disjoint bands, aggregating them for a single link.

Regarding telecommunication and cloud/edge services regulations and standards, the service providers must adhere to the existing ones unless new ones are created due to the development of new virtual network functions and O-RAN solutions. It is important to stress that virtualized networks with service-based architecture are vulnerable to hackers' attacks. Cybersecurity standards, the zero-trust principle, continuous and rigorous security practices and tests for software, hardware and user equipment, traffic monitoring, threat protection, and data privacy are issues that must be handled and included in the platform business model.

Finally, let us consider the components of this mentioned business model. In Table I, we provide some of these components for value creation and value delivery (part i) and ii) of the business model components mentioned above) together with analogies with Uber. There, selected analogies between Uber and uberized 2C networks are provided. (Note that it is not our intention to create a ready-to-apply business model for telco uberization.) We will address part iii) (value capturing) in the next section when we discuss the economy and pricing of 2C resources. The last row in Table I refers to legal regulations that are important elements of the business operating environment.

Having discussed the platform business model, the network architecture with distributed 2C resources, and Uber analogies for future 2C services, we believe an *Uber-like* business model can be potentially viable for owners of 2C resources. However, let us emphasize that Uber's exact business model is unlikely to be replicated by new players in this market. Due to the complexity and variety of today's networks, the diversity of engaged resources, the required security protocols, and telecom and spectrum regulations, their share of the common market may be narrower than that of car owners in Uber's market. A prospective player may share some of the resources in possession of this player but pay for other resources owned by another player. The uberized 2C market will most probably include complex sharing mechanisms.

## III. Uberized 2C resources

Let us consider the scenery of a 2C network with various customers and service providers. So far, cloud computing has been promoted as the data processing and storage technology for various applications. Recently, it has become apparent that this approach will not be able to withstand requirements of mentioned applications due to limited radio, network, and energy resources, high latency and packet errors in long-distance links, as well as additional control traffic on the mixed wireless and wired links in the cloud-based networks.

### A. Service and network scenario

Future 2C network design emphasizes edge computing as a viable option for many critical applications. A *fog network* has been promoted as a hierarchical, balanced network organization where 2C tasks can be performed locally [10]. Fog is an architecture for communication and computing, including information processing, storage, control, and networking, that distributes services closer to end users along the cloud-to-things continuum. It is more suited for 2C services than centralized (cloud-based) ones. It supports a growing variety of applications, including ultra-high reliability and low-latency communication (URLLC) services.

An illustrative example of the fog network is presented in Fig. 3. In the end-user/things tier, devices are connected, such as vehicles, smartphones, etc. In the cloud tier, powerful data servers are placed. In the fog tier, connected computing machines (PCs, computing clusters, etc.) capable of data processing, communication, and storage are located. The fog tier can have multiple hierarchical layers, and communication between them is vertical and horizontal. This architecture incorporates multiple types of communication links (wireless/wired, fronthaul/backhaul, optical/coaxial) and multiple types of computing devices.

Depending on the quality requirements expressed by Key Performance Indicators (KPIs) and the QoS metrics guaranteed for a particular 2C task, it can be executed locally or delegated to a (closer or more distant) fog node or the cloud. In Fig. 3, information flow is visualized for some example use-cases of data flow and computing tasks: vehicular communication (URLLC type), task offloading from a device with low computing power and limited memory, content caching (enhanced Mobile Broad Band communication – eMBB type), remote control (URLLC for medical or industrial applications) or telemetry data flow (massive Machine-Type Communication – mMTC). Note that QoS metrics should be guaranteed by a service provider and monitored by a customer, just like in contemporary networks and cloud servers. Customer experiences in this regard will shape the demand-supply structure and will impact the choice of a service provider. Note that the end-to-end quality provisioning should be also supported by regulations, which would impact all tiers of the fog, including homologation of the equipment, spectrum policy for wireless communication, software standards, security protocols, communication network and interoperability regulations, etc.

In the considered network, the 2C resources can be shared by parties owning (or licensing) them based on the platform business model. Moreover, the fog network concept can be used as a universal framework to distribute resources and services, and the *Uberization platform* from Fig. 3 as the *Platform stack* from Fig. 2 to manage, orchestrate, and secure the distributed resources and services.

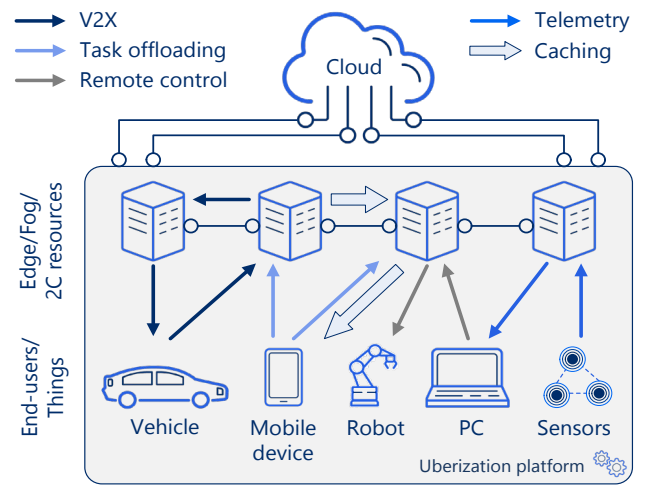

Fig. 3. Uberized 2C network and its resources

*B. Optimization of 2C resources*

The considered 2C network architecture allows for the optimization of the use of 2C resources. Naturally, there is a trade-off between processed data volume, delay, reliability, and energy consumption in such a network, e.g., computing machines in the fog nodes may consume more energy per operation than optimized cloud servers but provide lower latency. The high computational complexity of the processing requests favors processing in a cloud. Delay and power consumption caused by transmitting data through the core network are offset by the high processing speed and computational power efficiency. Conversely, requests which require relatively few operations per bit to process requests are best served by nearby edge devices or fog nodes. Fig. 4 presents example results of resource optimization for various scenarios of the cloud proximity in the fog network with parameters defined in [11].

There has been much research on computing task allocation optimization in fog networks, e.g., [11] and [12]. Most of them aim to minimize the energy cost of using 2C resources for required KPIs, e.g., latency, reliability, bit rate, etc. For the reason of computational complexity of the optimization algorithms themselves, there are simplifying assumptions made. This is natural since the inclusion of the wireless and wired transmission costs over the diverse network infrastructures, as well as the computing cost of using diverse machines, complicates the cost-benefit optimization grossly. First of all, usually, there needs to be complete information available on the energy consumption of all elements of networks, and second, complex optimization problems have to be implemented centrally and their results distributed to all network users, which may introduce additional delay and cost. A survey of multiple optimization algorithms proposed for fog networks is in [13].

The above issues could be bypassed in self-learning networks and systems that can autono-mously manage resources and control functions. It is impractical to transmit a massive amount of local data to the centralized cloud for training and inference. New neural network architectures and their associated communication-efficient training algorithms at the network edge are being developed. Such architectures also pose challenges: limited access to training data, low inference accuracy, lack of generalization, and limitations of processing power and memory for edge devices.

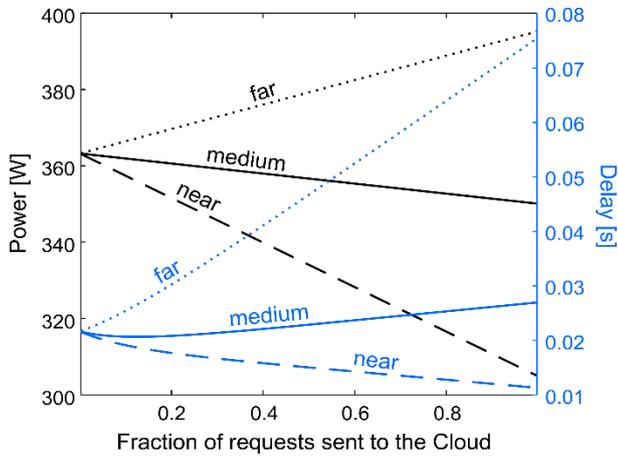

Fig. 4. Power consumption and delay for near (100 km), medium (2000 km) and far (8000 km) cloud distance.

*C. Cooperation and competition for 2C resources*

In the considered market, it is expected that there may be demand for 2C resources exceeding the supply. In such a case, competition for 2C resources among potential consumers will be observed. Conversely, when supply exceeds the demand, there may be competition between the suppliers of the resources. This is where game theory may come into play for analyzing non-cooperative and cooperative usage of resources in the fog (see several chapters on the topic in [10]).

A new approach to resource sharing is based on a methodology called coopetition (cooperative competition). It creates an added value in cooperation where value distribution is an element of competition. It is a business ideology taken from insights gained by game theory. The aim of coopetition is to move the market from a zero-sum winner-takes-all game to an environment where the result benefits everyone [14]. In [15], this methodology has been used in resource allocation in wireless networks. The competition phase applies the Cournot competition model, and the cooperation phase is the coalitional game. Due to the flexibility in the definition of each phase, the network may operate under different policies, supporting hierarchical traffic, fairness, and resource utilization efficiency. Coopetition fits the scenario where service providers or consumers may cooperate and compete in getting access or offering 2C resources.

*D. Fogonomics and resource pricing*

Recently, a new concept has been coined, namely *Fogonomics* (economics of the fog) [10]. Fogonomics deals with the economic factors that affect the design of fog architectures, the economy of resource sharing, interactions between consumers and service providers, and resource pricing. Heterogeneity inherent in fog architectures with uberized 2C resources leads to the following pricing challenges: (i) the coexistence of heterogeneous networks that requires choosing between the interfaces used and multiple network providers, (ii) scaling up and down the computing needs and choosing the place of computing by a variety of end devices, and (iii) accordance to performance metrics of heterogeneous services offered [10].

Regarding network/communication resource pricing, today's mobile Internet service providers (IPSs) usually offer capped and usage-based data plans that charge a base payment for a limited amount of data and degrade QoS or charge steep overage costs above this amount. A uberized fog network can alleviate growing network congestion by dynamically pooling resources and conducting calculations on local devices. This will decrease the quantity of data that needs to be transferred, lowering the cost of Internet access. However, finding the appropriate pricing for various QoS requirements is a challenging research problem since the devices have multiple options for network connectivity with diverse network interfaces. Virtual IPSs with uberized services may attract consumers since they ensure access to multiple network interfaces.

Regarding computing resources nowadays, the cost of CPU, memory or storage access makes up most payments. The computational tasks can be performed using serverless functions upon request or on dedicated instances with predefined configurations. Fog applications will incur various financial and nonfinancial costs for using resources on various devices. Thus, cost optimization services will be needed to choose the best application configuration for cost and performance.

In a heterogeneous 2C scenario, individual devices can be selected to play various roles (gateways, computing nodes, etc.). This flexibility and the trade-offs between offered pertinent QoS metrics (e.g., energy efficiency vs. latency vs. reliability and security) should be considered when deciding 2C resource pricing. For the aforementioned URLLC applications, for instance, latency and reliability are crucial. The case studies for resource pricing in such a scenario can be found in [10].

Pricing schemes for uberized 2C resources should attract a higher market share and shape user demand. These may require surge pricing to ensure the trustworthiness of devices offering resources and sharing-incentive mechanisms. They may include pay-per-use pricing, optimal taxing for using the network, auction-based pricing, or volume-discount pricing. Pricing models like these can significantly contribute to user satisfaction and possibly shape demand structure to reduce congestion. When designed properly, they can generate higher revenue for service providers.

## IV. Conclusions

Uberization has already become a socio-economic phenomenon. We believe that telcos and companies offering computing services could be at the forefront of deploying those technologies across their infrastructure and developing innovative offerings that disrupt their prevailing products.

Recent progress in fog network architectures, optimization, and management of 2C resources, as well as in the practice and economics of their sharing allows considering uberization as a promising direction of telco business democratization. Given the 2C market opening constraints, the roadmap for its implementation includes the following milestones: i) further progress in networks virtualization (including RANs), network slicing and supporting mechanisms (including security procedures), ii) advancement in spectrum sharing mechanisms and policies iii) advancement in 2C resource pricing algorithms taking QoS into account, and iv) legal regulations and resolutions supporting open competition, rights, and obligations of the service providers and customers.

Note that concerns on the quality of service offered and the employment terms in platform businesses are nowadays subject to various resolutions, e.g., under EU policies (see "EU rules on platform work"), and should be dispelled in a near future. However, provisioning of the end-to-end quality of 2C services is a complex issue that requires the definition of 2C quality metrics, and the study of methods that guarantee them in diverse market scenarios with multiple service providers.

**Hanna Bogucka** is a full professor and the Director of the Institute of Radiocommunications at Poznan University of Technology (PUT). She is also a co-founder of RIMEDO Labs. Her research focus on cognitive and green radio. She is a member of the Polish Academy of Sciences. She serves the IEEE ComSoc Fog/Edge Industry Community as European Chair and the IEEE ComSoc Board of Governors as EMEA region representative.

**Bartosz Kopras** is a PhD student at the Institute of Radiocommunications at PUT. His area of interest is communication and computing and in particular, energy-efficient communication and computing in heterogeneous fog networks.